\documentclass[aps,prb,reprint,showpacs,superscriptaddress]{revtex4-1}
\usepackage{amssymb,amsmath,amsbsy,graphicx,hyperref,xcolor,csquotes,braket}
\DeclareMathOperator{\sgn}{sgn}

\begin{document}
\title{Chiral properties of topological-state loops}

\author{Marko M. Gruji\'c}\email{marko.grujic@etf.bg.ac.rs}
\affiliation{School of Electrical Engineering, University of Belgrade, P.O. Box
3554, 11120 Belgrade, Serbia} \affiliation{Department of Physics, University of
Antwerp, Groenenborgerlaan 171, B-2020 Antwerp, Belgium}
\author{Milan \v{Z}. Tadi\'c}
\affiliation{School of Electrical Engineering, University of Belgrade, P.O. Box
3554, 11120 Belgrade, Serbia}
\author{Fran\c{c}ois M. Peeters}
\affiliation{Department of Physics, University of Antwerp, Groenenborgerlaan
171, B-2020 Antwerp, Belgium}

\begin{abstract}
The angular momentum quantization of chiral gapless modes confined to a
circularly shaped interface between two different topological phases is
investigated. By examining several different setups, we show analytically that
the angular momentum of the topological modes exhibits a highly chiral
behavior, and can be coupled to spin and/or valley degrees of freedom,
reflecting the nature of the interface states. A simple general 1D model, valid
for arbitrarily shaped loops, is shown to predict the corresponding energies
and the magnetic moments. These loops can be viewed as building blocks for
artificial magnets with tunable and highly diverse properties.
\end{abstract}
 \maketitle

\section{Introduction}\label{I}

The appearance of gapless modes at the interface between insulating phases with
distinct topologies is one of their quintessential traits, and holds many
promises for novel device designs.\cite{martin08,qiao11,qian14,pan14} This
feature is usually attributed to the so-called bulk-edge correspondence.
Crudely speaking, in order to \enquote{unwind} one band structure into another
one must close the gap at some point, rendering protected localized states,
that usually display chiral properties.\cite{qi11} The prime example of this is
the Quantum spin Hall (QSH) effect. There, counter-propagating modes belonging
to opposite spins will emerge within the bulk band gap at the boundary between
two time-reversal symmetric insulators with different $Z_2$ topological
invariants.\cite{kane05a,kane05b,bernevig06}

While the manifestation of chirality in electrical transport has been
thoroughly
researched,\cite{zarenia11,zarenia12,tudor12,jung12,qiao14,kim14,abergel14,wang14,rachel14}
in this paper we will examine in detail what happens when the chiral states are
forced into a loop delineating two topologically distinct insulating domains.
This is somewhat different and more general than the case studied in Ref.
\onlinecite{shan11}, where the impact of circular vacuum domains in $Z_2$
topological insulators on transport was studied. In particular we will show
that, depending on the particular topology and the type of material on either
side of the loop, the chirality manifests as a spin and/or valley coupling
between the states bound to the orbit and their angular momentum, making these
loops potential building blocks for hybrid magnetic structures. More
importantly, we will also show that a general 1D model captures all of the
physics involved, even for noncircular loops, and discuss the limited impact of
intervalley scattering. Our work is in particular focused on group IV honeycomb
monolayers, which we analyze within a continuum approximation by employing the
Dirac equation and the $\pi$-band tight-binding (TB) model.

The paper is organized as follows. In section \ref{II} the analytical solutions
for circular interfaces are studied using the Dirac equation. In section
\ref{III}, the tight-binding method is used to examine the impact of the
magnetic fields on the topological states for varying loop geometries, as well
as to establish the validity of a general formula for energy levels and
magnetic moments of the states bound to the loops. Finally, in section \ref{IV}
we discuss the practical consequences of our proposal and summarize our
results.

\section{The continuum approach: circular interfaces}\label{II}

We start by introducing our general model, which is the Dirac equation valid
for honeycomb-lattice materials
\begin{equation}\label{kane-mele-haldane}
H=\hbar v_F\left(\tau k_x\sigma_x+k_y\sigma_y\right)+s\tau\Delta_{KM}\sigma_z+\tau\Delta_H\sigma_z+\Delta_{SP}\sigma_z.
\end{equation}
Here $\tau=+1$ ($\tau=-1$) denotes the $K$ ($K^{\prime}$) valley and $s=+1$
($s=-1$) labels spin up (spin down). $\Delta_{KM}$ is Kane and Mele's
phenomenological term modeling spin-orbit coupling,\cite{kane05b} while
$\Delta_H$ was introduced by Haldane as a toy model for a non-trivial band
structure in the absence of magnetic field,\cite{haldane88} inducing the
quantum anomalous Hall (QAH) phase. Finally, $\Delta_{SP}$ is the staggered
potential breaking the inversion symmetry, and opening a trivial band gap
unlike the previous two.

\begin{figure}
\centering
\includegraphics[width=8.6cm]{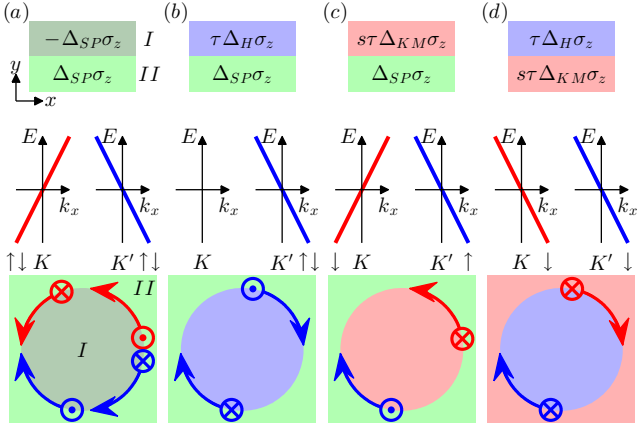}
\caption{(a) - (d) The considered set of interfaces between insulating phases.
The upper row shows the parameters related to each region,
while the middle row depicts the corresponding dispersion of the emerging modes.
The bottom row illustrates the hypothesis for the persistent currents in the closed loops
inferred from the linear dispersion relations. Red (blue) lines depict the states belonging
to the $K$ ($K^{\prime}$) valley, while the spin is denoted by the projection of the arrows.}
\label{fig1}
\end{figure}

An interface between any of the aforementioned insulating phases will support
boundary modes. This can be shown by solving Eq. \eqref{kane-mele-haldane} for
the geometry depicted in the upper row of Fig. \ref{fig1}. Note that in the
domains $I$ and $II$ only a single mass term of the form
$\Delta_{I/II}\sigma_z$ is used in Eq. \eqref{kane-mele-haldane}. For instance,
in case (c) $\Delta_I=s\tau\Delta_{KM}$ and $\Delta_{II}=\Delta_{SP}$. Then the
evanescent modes, decaying exponentially away from the interface read
\begin{equation}
\Psi_{I/II}=e^{ik_xx}e^{\mp \kappa_{yI/II}y}\left[\begin{array}{c}1\\\frac{\hbar v_f\left(\tau k_x\mp \kappa_{yI/II}\right)}{\Delta_{I/II}+E}\end{array}\right],
\end{equation}
and we are interested in solutions $E=\pm\hbar v_Fk_x$ with
$\kappa_{yI/II}=\left|\Delta_{I/II}\right|/\hbar v_F$. Matching the
wavefunctions at the boundary leads to the dispersion relations shown in the
middle row of Fig. \ref{fig1}. These demonstrate chirality of the topological
modes bound to the interface.\cite{qiao11,kane05b} Unlike the cases in Figs.
\ref{fig1}(c) and (d), the first two cases are spin degenerate.

\begin{table}
\caption{Summary of the numerical solutions. The symbol \enquote{$\times$}
means that there are no allowed solutions, while the allowed modes are
characterized by the corresponding angular momentum numbers $j$.}
    \begin{tabular}{| l | c | c | c | c |}
    \hline
     & $\uparrow K$ & $\uparrow K^{\prime}$ & $\downarrow K$ & $\downarrow K^{\prime}$ \\ \hline
    (a) & $j>0$ & $j<0$ & $j>0$ & $j<0$ \\ \hline
    (b) & $\times$ & $j<0$ & $\times$ & $j<0$ \\ \hline
    (c) & $\times$ & $j<0$ & $j>0$ & $\times$ \\ \hline
    (d) & $\times$ & $\times$ & $j<0$ & $j<0$ \\
    \hline
    \end{tabular}
\label{tabela}
\end{table}

We now present the main premise of our paper. Instead of straight interfaces,
we consider circular ones, depicted in the bottom row of Fig. \ref{fig1}. The
chirality of the boundary modes suggests that the angular momentum of the
states in these loops must obey certain selection criteria depending on the
particular phases in regions $I$ and $II$. For instance, in the cases (b) and
(d) one would expect to see that only negative angular momentum valley and spin
filtered states are allowed, respectively

To explore this explicitly, we solve the problem in polar coordinates, in which
the Hamiltonian \eqref{kane-mele-haldane} reads
\begin{equation}\label{kane-mele-haldane-radial}
H=\left[\begin{array}{cc}\Delta_{I/II} & \pi^-\\
\pi^+ & -\Delta_{I/II}\end{array}\right],
\end{equation}
where $\pi^{\pm}=\tau\hbar v_Fe^{\pm
i\tau\phi}\left(-i\partial_r\pm\frac{\tau}{r}\partial_\phi\right)$. Because of
the radial symmetry the total angular momentum
$J_z=L_z+\tau\frac{\hbar}{2}\sigma_z$ is a good quantum number denoted by
$j$,\cite{recher07,grujic11} and therefore the solution can be sought in the
form
\begin{equation}
\Psi=e^{i(j-\tau/2)\phi}\left[\begin{array}{c}\chi_A(r)\\e^{i\tau\phi}\chi_B(r)\end{array}\right].
\end{equation}
The coupled system of equations reduces to the differential equation
\begin{equation}
\left[r^2\frac{d^2}{dr^2}+r\frac{d}{dr}-\frac{\Delta_{I/II}^2-E^2}{\hbar^2v_F^2}r^2-(j-\tau/2)^2\right]\chi_A=0.
\end{equation}
Having in mind that we are interested in states lying within the bulk band gap
$\Delta_{I/II}^2-E^2>0$, so that the solutions are the modified Bessel
functions $I_{j-\tau/2}(r)$ and $K_{j-\tau/2}(r)$, which are exponentially
divergent at $r\rightarrow\infty$ and $r=0$, respectively. The radially bound
modes read
\begin{equation}
\Psi_I=e^{i(j-\tau/2)\phi}\left[\begin{array}{c}I_{j-\tau/2}\left(z_I(r)\right)\\-i\tau e^{i\tau\phi}\frac{\sqrt{\Delta_I^2-E^2}}{\Delta_I+E}I_{j+\tau/2}\left(z_I(r)\right)\end{array}\right],
\end{equation}
and
\begin{equation}
\Psi_{II}=e^{i(j-\tau/2)\phi}\left[\begin{array}{c}K_{j-\tau/2}\left(z_{II}(r)\right)\\i\tau e^{i\tau\phi}\frac{\sqrt{\Delta_{II}^2-E^2}}{\Delta_{II}+E}K_{j+\tau/2}\left(z_{II}(r)\right)\end{array}\right],
\end{equation}
where $z_{I/II}=\frac{\sqrt{\Delta_{I/II}^2-E^2}}{\hbar v_F}r$. Matching
$\Psi_I$ and $\Psi_{II}$ at the interface $r=R$ results in the eigenvalue
equation
\begin{align}\label{tranjna}
\frac{\sqrt{\Delta_{II}^2-E^2}}{\Delta_{II}+E}K_{j+\tau/2}\left(z_{II}(R)\right)I_{j-\tau/2}\left(z_I(R)\right)\nonumber\\
+\frac{\sqrt{\Delta_I^2-E^2}}{\Delta_I+E}I_{j+\tau/2}\left(z_I(R)\right)K_{j-\tau/2}\left(z_{II}(R)\right)=0,
\end{align}
which we solve numerically.

Our numerical results are summarized in Table \ref{tabela} for four cases shown
in Fig. \ref{fig1}. For those spin and valley flavors for which there exists a
solution for a given interface we give the allowed angular momentum values,
while the cases for which no solution is found are indicated by
\enquote{$\times$}. Only results for positive energies are given, since the
negative energy solutions are the same, albeit with opposite values of $j$. By
inspecting Table \ref{tabela} and comparing it with Fig. \ref{fig1} one can see
that the numerical results agree with the expectations deduced from the
chirality of the topological modes. Namely, the angular momentum of the states
bound to the loops not only display the expected orientation but are also: (a)
valley-coupled, (b) valley-filtered, (c) spin-valley-filtered, and (d)
spin-filtered.

\begin{figure}
\centering
\includegraphics[width=8.6cm]{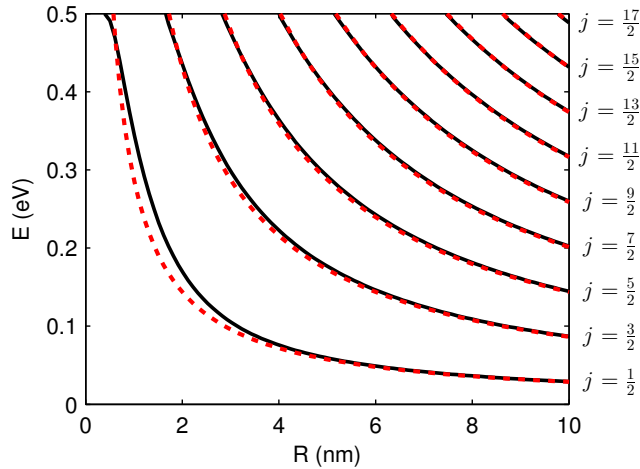}
\caption{Variation of the numerically computed energy levels
with radius $R$ (solid black lines), and the results from semiclassical
quantization (dashed red lines) for the loop considered in Fig. \ref{fig1}(a).
Here $\Delta_{SP}=0.5$ eV.}
\label{fig2}
\end{figure}

In Fig. \ref{fig2} the solid black lines are the numerically computed energies
of the bound states for the radial interface corresponding to Fig.
\ref{fig1}(a) for spin up in the $K$ valley. Note that for all the modes that
are allowed, the set of energies is the same (only the angular momentum number
$j$ varies), such that it suffices to show the spectrum for only one particular
case. One can see that the larger loops can support states with larger $j$.
Note also that, as a rule, there exists only one state for a given $j$. This is
due to the exponential localization at the loop, which prevents oscillatory
behavior in the radial direction and the emergence of a corresponding degree of
freedom.

Interestingly, the energy curves follow a $1/R$ behavior with very high
precision, a property also seen in states bound to the holes punched through
topological insulators.\cite{shan11} It is relatively easy to obtain a good
analytical fit of the energy levels using the following reasoning. Given the
exponential localization of the modes near $r=R$, the loops effectively become
1D closed lines carrying states with pseudospin degree of freedom, described by
$\psi$. Then, by requiring single-valuedness of $\psi$ upon one full revolution
we must have $\psi(\phi+2\pi)=-e^{i\theta}\psi(\phi)$, where the contour is
oriented counterclockwise. The minus sign comes from the spinorial nature of
the state, and it forces the phase factor accumulated along the trajectory
$\theta=\sgn(j)pL/\hbar$ ($L=2\pi R$ is the loop length) to be quantized in
half integer multiples of $2\pi$. Denoting this half integer number by $j$, and
recalling the dispersion of the topological modes $E=v_F p$, yields the
following expression
\begin{equation}\label{energija}
E=\left|j\right|\frac{\hbar v_F}{R},
\end{equation}
for the energy levels. They are shown in Fig. \ref{fig2} by dashed red lines,
and are in excellent agreement with the numerical results. Only small
deviations are found for loops with small radius. This is to be expected, given
that for small loops overlap of the wavefunction must occur, permitting
tunneling of the modes between sections of the loop.

\section{The tight-binding method: generalization for arbitrarily shaped loops}\label{III}

Having analytically elucidated the chiral nature of the angular momenta, we now
employ the TB method to study the behavior of the loops in a magnetic field. On
the one hand, this is important because it can offer insights into intervalley
scattering. Indeed, the most critical case is the setup (a), since valley
flipping can cause back-reflection and mixing of the states, something not
captured within the continuum approach. All the other loops are more robust,
since either there are no counter-propagating states ((b) and (d)), or
backscattering requires the simultaneous flipping of the valley and spin
indices (setup (c)). On the other hand, the TB method will help us to resolve
important details of the magnetic moments originating from the loops, even for
noncircular geometries. The TB Hamiltonian that we employed reads
\begin{equation}\label{eq:Hamiltonian}
H=-t\sum_{\langle n,m\rangle}e^{i\varphi_{nm}}c^{\dagger}_{n}c_{m}+
i\frac{s\Delta_{KM}}{3\sqrt{3}}\sum_{\langle\langle n,m\rangle\rangle}\nu_{nm}e^{i\varphi_{nm}}c^{\dagger}_{n}c_{m}.
\end{equation}
We use $t=2.7$ eV for the hopping between nearest neighbor $p_z$ orbitals,
while the second term is the SOC. Note that $\nu_{nm}=+1$ ($\nu_{nm}=-1$) if an
electron makes a right (left) turn at the intermediate atom when hopping from
site $m$ to site $n$. The Peierls term
$\varphi_{nm}=-\frac{e}{\hbar}\int_{\mathbf{r_m}}^{\mathbf{r_n}}\mathbf{A}\cdot
d\mathbf{l}$ accounts for the phase the electron acquires while traveling in
the presence of the magnetic field. The staggered potential is simply added
along the main diagonal of the Hamiltonian, like in the continuum model.

\begin{figure}
\centering
\includegraphics[width=8.6cm]{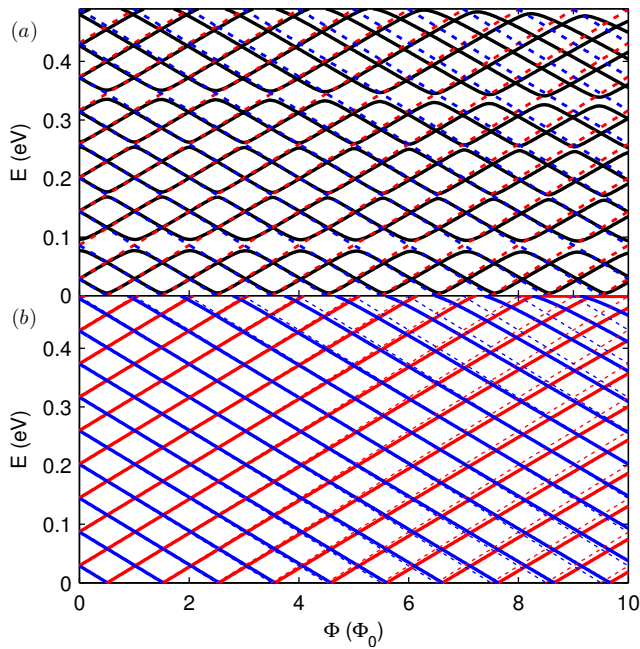}
\caption{The magnetic field dependence of the energy spectrum for the loops depicted
in (a) Fig. \ref{fig1}(a) and (b) Fig. \ref{fig1}(c). We used
$\Delta_{SP}=\Delta_{KM}=0.5$ eV. Solid curves
depict the TB results, while the dashed lines are given by Eq. \eqref{energijeodb}.
Red (blue) depicts the states belonging to the $K$ ($K^{\prime}$) valley.}
\label{fig3}
\end{figure}

Our calculations were performed for a loop of radius $R=10$ nm contained within
a larger circular quantum dot, with a $15$ nm radius, so that a proper decay of
the topological modes is ensured. The results are shown by solid lines in Figs.
\ref{fig3}(a) and (b) for the loop setups depicted in Figs. \ref{fig1}(a) and
(c), respectively, as a function of the magnetic flux through the loop
$\Phi=BR^2\pi$ in units of the flux quantum $\Phi_0=h/e$. For setup (a) the
results are shown in black, since the TB model cannot resolve the two valleys,
while for setup (c) (Fig. \ref{fig3}(b)) the valleys can be distinguished due
to the additional spin polarization. In Fig. \ref{fig3}(a), for $B=0$ the
majority of the levels are degenerate indicating the preservation of the valley
index. Only a handful of levels (for instance those around $0.1$ eV) are offset
by an equally small amount in the opposite directions with respect to the
continuum solutions. This behavior indicates that such states are susceptible
to intervalley scattering, causing bonding and anti-bonding states. As the
magnetic field is turned on, the remaining valley degeneracy is lifted due to
the opposite magnetic moments (see below). Moreover, an anticrossing behavior
occurs at energies where intervalley scattering is prominent, resulting in an
Aharonov-Bohm-like pattern of the curves. On the other hand, for the setup (c),
such scattering is impossible in the TB picture due to the absence of
spin-flipping terms in the Hamiltonian, as demonstrated by Fig. \ref{fig3}(b).
Consequently, the spin up in the $K^{\prime}$ valley (solid blue curve), with a
magnetic moment oriented along the magnetic field, and the spin down in the $K$
valley (solid red curve) with the opposite orientation of the magnetic moment
are independent of each other, and experience a rigid, linear, decrease and an
increase in energy, respectively.

In order to capture the behavior of the topological modes in the presence of a
magnetic field, we again consider the phase change upon a single rotation
around the loop. Now, due to the magnetic field an additional phase factor of
the form (assuming $B$ is constant inside the loop)
$-\frac{e}{\hbar}\oint\mathbf{A}\cdot d\mathbf{l}=-\frac{e}{\hbar}BS$ appears,
where $S$ is the loop area. As before, $\theta$ must be a half integer multiple
of $2\pi$, yielding
\begin{equation}\label{energijeodb}
E=\left|j\right|\frac{hv_F}{L}+\sgn(j)\frac{ev_F}{L}SB.
\end{equation}
These levels are depicted by dashed curves in Fig. \ref{fig3} and show
excellent agreement with the TB results, with small deviations at very large
magnetic fields.\cite{latyshev14} Note that the change in energy due to
magnetic field can be captured entirely by the classical formula for the energy
of a closed current contour given by $-\mu B$. Besides the area of the loop,
the magnetic moment of a classical contour $\mu=IS$ depends on the current due
to one topological mode $I=\sgn(j)(-e)/T$, where $T=\sgn(j)(-e)v_F/L$ is the
period, and matches the one in expression Eq. \eqref{energijeodb} exactly.

\begin{figure}
\centering
\includegraphics[width=8.6cm]{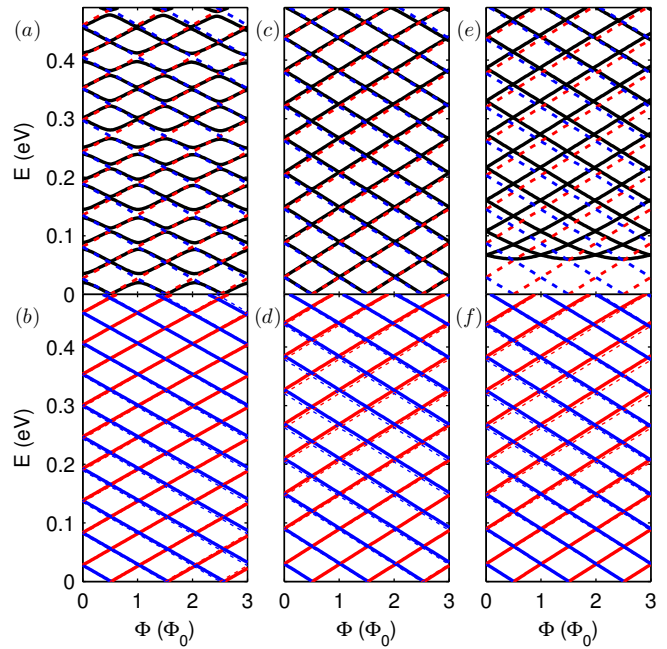}
\caption{The same as in Fig. \ref{fig3}, but for an elliptic loop ((a) and (b)),
hexagonal zigzag loop ((c) and (d)), and hexagonal armchair loop ((e) and (f)).}
\label{fig4}
\end{figure}

Most importantly, note that Eq. \eqref{energijeodb} is written in a general
form, without specifying the precise shape of the loop, since it depends only
on its length and area. In fact, we have verified numerically by using the TB
method that the above formula accurately predicts the energies and magnetic
moments of the states in loops with arbitrary shape. In particular, in Fig
\ref{fig4} we show the results for the same two setups as in Fig. \ref{fig3},
but for a elliptic, hexagonal zigzag and hexagonal armchair loops. Note that
$j$ does not denote the total angular momentum now, due to the lack of
rotational symmetry. However, it is still required to be a half-integer, while
its sign reflects the orientation of the chiral current. For the elliptic loop
(Figs. \ref{fig4}(a) and (b)) we have used $R_1=5$ nm, $R_2=15$ nm,
$S=R_1R_2\pi$, and for the loop length we have used the approximation of the
form
\begin{equation}
L=\pi(R_1+R_2)\left(1+\frac{3d}{10+\sqrt{4-3d}}\right),
\end{equation}
where $d=\left(R_1-R_2\right)^2/\left(R_1+R_2\right)^2$. The hexagonal loops
are distinguished by the number of zigzag ($N_{ZG}$) or armchair ($N_{AC}$)
segments along one edge. The loop length ($L=6l$) and area
($S=6l^2\sqrt{3}/4$), depend on the length of one side of the hexagon
$l_{ZG}=N_{ZG}\sqrt{3}a$ ($l_{AC}=(3N_{AC}-2)a$), where $a$ is the nearest
neighbor distance. The results depicted in Figs. \ref{fig4}(c) and (d), and (e)
and (f) are for $N_{ZG}=42$ and $N_{AC}=25$, respectively. All the loops are
surrounded by a region large enough to ensure the decay of the topological
modes.

On the one hand, for setup (c) (where counterpropagating states belong to
opposite spins, lower panels), it can be seen that regardless of the shape, the
persistent current is robust, and that Eq. \eqref{energijeodb} predicts the
energy levels with great accuracy. The very small mismatch that appears we
attribute to the fact that the sharp turns and corners can effectively reduce
the loop length and area due to tunneling, and hence cause some deviation from
the expected behavior. On the other hand, the spectrum of setup (a) (upper
panels) shows that intervalley scattering is geometry dependent, as one would
expect, although Eq. \eqref{energijeodb} still provides a very good
approximation for the results in general. While elliptic loops display stronger
intervalley scattering and subsequent hybridization than the circular loops,
intervalley scattering in zigzag hexagonal loops is practically nonexistent.
Finally, in armchair hexagonal loops a gap is opened for low-energy chiral
states, a situation resembling the appearance of mass barriers in graphene
quantum rings.\cite{roman13} Remarkably, away from this energy region a
behavior indicating chirality preservation is observed, although the valley
degree of freedom is ill-defined in this structure. Therefore, to a good
approximation one can say that the electronic and magnetic properties of these
loops are largely independent on the precise details of the edges near the
interface. This is unlike the case in regular graphene nanorings, where
geometry and boundary of the rings can hinder the fidelity of the simplified 1D
models,\cite{costa14} with the main reason for this contrast being the
exponential localization of the topological states.

\section{Discussion and the conclusion}\label{IV}

Now we consider the possible applications of these effects. By controlling the
valley population separately in the setup (a) for example (in loops with well
preserved valley index), one could in turn control the total magnetization of
the loop, since all orbits within a given valley have the same orientation of
the persistent current. In setup (b) the $K$ valley does not host topological
boundary modes, unlike the $K^{\prime}$ valley, where the persistent current
has the same orientation regardless of spin. Hence, these particular loops
should be inherently magnetic. Alternatively, in setup (c) one can control the
total magnetization by adjusting the spin or spin-valley
populations.\cite{grujic14a} Finally the loops in setup (d) not only display
nonzero magnetism, as those in setup (b), they have the additional benefit that
the allowed topological modes have their spins aligned, which can contribute to
the overall magnetic moment. Therefore, these loops have a tremendous potential
for the design of novel magnetic structures and devices. One can imagine an
array of these loops laid out in a pattern designed to suit a particular
application. Moreover, since the states racing along these loops are
exponentially localized in the radial direction the array can be densely
packed, enabling the creation of hybrid magnets.

Before we conclude, let us reflect on the practical issues concerning the real
materials that are required to create these structures. On the one hand, it is
well known that inversion symmetry can be broken by layering graphene with
boron-nitride, which has a naturally occurring staggered potential since it's
sublattices are composed out of different atomic species.\cite{hunt13}
Moreover, in-plane heterostructures of graphene and boron-nitride have been
demonstrated in recent experiments.\cite{liu13,liu14} In particular, in Ref.
\onlinecite{liu13} the authors presented a matrix array of circular graphene
dots embedded in boron-nitride. Since the resulting structure was transferable,
one can imagine layering it with a substrate which would enhance SOC in
graphene, such as ${\rm WS}_2$ for instance,\cite{avsar14} thereby creating a
macroscopic pattern of chiral closed-contour currents of the type shown in Fig.
\ref{fig1}(c).

On the other hand, given the buckled lattice structure of silicene, germanene
and stanene,\cite{cahangirov09,liu11,xu13} breaking of the inversion symmetry
can be realized by external gates, while relatively heavy constituent atoms
ensure nonzero SOC. Patterning gates on top of them could likewise induce
chiral persistent currents described in this paper, making them detectable for
instance by magnetic force microscopy. Additionally, there are theoretical
proposals for inducing the QAH effect in graphene and silicene, although
experimental validation is still
lacking.\cite{kitagawa11,qiao14b,barnea12,wright13,zhang13,kaloni14}

As we have shown, the larger the bulk band gap and the loop radius, the larger
the magnetization will be. However, large loops and inherent spin and valley
relaxation mechanisms will inevitably lead to scattering, spoiling the effect
to a certain degree. Nevertheless, no amount of scattering can change the
orientation of the persistent current, which is related to the topology of the
surrounding band structures, and the emptied states should quickly become
repopulated. Note that, while the interplay of orbital magnetism emerging in
insulating Dirac systems,\cite{xiao07,zhang11,grujic14b} and the magnetism of
topological-state loops remains to be explored, the main obstacle for realizing
this effect is most likely related to the size of the bulk band gaps. This can
not only restrict the amount of persistent current supported by the loops, but
more importantly limit the temperature range of applicability.

To summarize, we investigated the behavior of electronic states bound to looped
interfaces between insulating phases of distinct topologies. These structures
can be a source of magnetic moments due to persistent currents flowing along
the interfaces. The chirality of the states bound to these loops manifests as a
chirality in the angular momentum and the corresponding magnetic moments. We
have shown that a simple 1D model captures both the qualitative and the
quantitative behavior of the electrons. By calculating energy levels and
magnetic moments in elliptic, hexagonal zigzag, and hexagonal armchair loops,
it was shown that the model's validity extends to noncircular loops as well.
Because of the intimate link between the magnetic moments and spin and valley
degrees of freedom, novel magnetic structures and devices can be envisioned.

\begin{acknowledgments}This work was supported by the Serbian Ministry of Education, Science and Technological Development, and the Flemish Science Foundation (FWO-Vl).
\end{acknowledgments}


\begin{thebibliography}{00}
\bibitem{martin08}I. Martin, Y. M. Blater and A. F. Morpurgo, Phys. Rev. Lett.
    {\bf100}, 036804 (2008).
\bibitem{qiao11}Z. Qiao, J. Jung, Q. Niu, and A. H. MacDonald, Nano Lett.
    {\bf11}, 3453 (2011).
\bibitem{qian14}Z. Qian, J. Liu, L. Fu and J. Li, Science
    {\bf346}, 1344 (2014).
\bibitem{pan14}H. Pan, X. Li, F. Zhang, and S. A. Yang, arXiv:1501.00114v1.
\bibitem{qi11}X.-L. Qi and S.-C. Zhang, Rev. Mod. Phys. {\bf83}, 1057 (2011).
\bibitem{kane05a}C. L. Kane and E. J. Mele, Phys. Rev. Lett. {\bf95}, 146802
    (2005).
\bibitem{kane05b}C. L. Kane and E. J. Mele, Phys. Rev. Lett. {\bf95}, 226801
    (2005).
\bibitem{bernevig06}B. A. Bernevig and S.-C. Zhang, Phys. Rev. Lett.
    {\bf96}, 106802 (2006).
\bibitem{zarenia11}M. Zarenia, J. M. Pereira Jr., G. A. Farias, and F. M.
    Peeters, Phys. Rev. B {\bf84}, 125451 (2011).
\bibitem{zarenia12}M. Zarenia, O. Leenaerts, B. Partoens, and F. M. Peeters,
    Phys. Rev. B {\bf86}, 085451 (2012).
\bibitem{tudor12}T. Tudorovskiy and M. I. Katsnelson, Phys. Rev. B {\bf86},
    045419 (2012).
\bibitem{jung12}J. Jung, Z. Qiao, Q. Niu, and A. H. MacDonald, Nano Lett.
    {\bf12}, 2936 (2012).
\bibitem{qiao14}Z. Qiao, J. Jung, C. Lin, Y. Ren, A. H. MacDonald, and Q.
    Niu, Phys. Rev. Lett. {\bf112}, 206601 (2014).
\bibitem{kim14}Y. Kim, K. Choi, J. Ihm, and H. Jin, Phys. Rev. B {\bf89},
    085429 (2014).
\bibitem{abergel14}D. S. L. Abergel, J. M. Edge, and A. V. Balatsky, New J.
    Phys. {\bf16}, 065012 (2014).
\bibitem{wang14}S. K. Wang, J. Wang, and K. S. Chan, New J. Phys. {\bf16},
    045015 (2014).
\bibitem{rachel14}S. Rachel and M. Ezawa, Phys. Rev. B {\bf89}, 195303 (2014).
\bibitem{shan11}W.-Y. Shan, J. Lu, H.-Z. Lu, and S.-Q. Shen, Phys. Rev. B
    {\bf84}, 035307 (2011).
\bibitem{haldane88}F. D. M. Haldane, Phys. Rev. Lett. {\bf61}, 2015 (1988).
\bibitem{recher07}P. Recher, B. Trauzettel, A. Rycerz, Y. M. Blanter, C. W. J.
    Beenakker, and A. F. Morpurgo, Phys. Rev. B {\bf76}, 235404 (2007).
\bibitem{grujic11}M. Gruji\'{c}, M. Zarenia, A. Chaves, M. Tadi\'{c}, G. A.
    Farias, and F. M. Peeters, Phys. Rev. B {\bf84}, 205441 (2011).
\bibitem{latyshev14}Non-topological Tamm-Dirac edge states, appearing near
    graphene nanoholes, also give rise to these levels, impacting the
    magnetotransport measurements, see: Yu I. Latyshev, A. P. Orlov, V. A.
    Volkov, V. V. Enaldiev,
    I. V. Zagorodnev, O. F. Vyvenko, Yu V. Petrov, and P. Monceau, Nat. Commun.
    {\bf4}, 7578 (2014).
\bibitem{roman13}I. Romanovsky, C. Yannouleas, and U. Landman, Phys. Rev. B {\bf87},
    165431 (2013).
\bibitem{costa14}D. R. da Costa, A. Chaves, M. Zarenia, J. M. Pereira Jr., G.
    A. Farias, and F. M. Peeters, Phys. Rev. B {\bf89}, 075418 (2014).
\bibitem{grujic14a}M. M. Gruji\'{c}, M. \v{Z} Tadi\'{c}, and F. M. Peeters,
    Phys.
    Rev. Lett. {\bf113}, 046601 (2014).
\bibitem{hunt13}B. Hunt, J. D. Sanchez-Yamagishi, A. F. Young, M. Yankowitz, B.
    J. LeRoy, K. Watanabe, T.
    Taniguchi, P. Moon, M. Koshino, P. Jarillo-Herrero, and R. C. Ashoori, Science {\bf340}, 1427 (2013).
\bibitem{liu13}Z. Liu, L. Ma, G. Shi, W. Zhou, Y. Gong, S. Lei, X. Yang,
    J. Zhang, J. Yu, K. P. Hackenberg, A. Babakhani, J.-C. Idrobo, R. Vajtai, J. Lou, and
    P. M. Ajayan, Nature Nanotech. {\bf8}, 119 (2013).
\bibitem{liu14}L. Liu, J. Park, D. A. Siegel, K. F. McCarty, K. W. Clark, W.
    Deng, L Basile, J.-C. Idrobo, A.-P. Li, and G. Gu, Science {\bf343}, 163
    (2014).
\bibitem{avsar14}A. Avsar, J. Y. Tan, T. Taychatanapat, J. Balakrishnan, G.K.W.
    Koon, Y. Yeo, J. Lahiri, A. Carvalho, A. S. Rodin, E.C.T. O'Farrell, G.
    Eda, A. H. Castro Neto, and B. \:{O}zyilmaz, Nat. Commun. {\bf5}, 4875
    (2014).
\bibitem{cahangirov09}S. Cahangirov, M. Topsakal, E. Akt\"{u}rk, H. \c{S}ahin,
    and S. Ciraci, Phys. Rev. Lett. {\bf102}, 236804 (2009).
\bibitem{liu11}Cheng-Cheng Liu, Wanxiang Feng, and Yugui Yao, Phys. Rev. Lett.
    {\bf107}, 076802 (2011).
\bibitem{xu13}Y. Xu, B. Yan, H. J. Zhang, J. Wang, G. Xu,
    P. Tang, W. Duan, and S. C. Zhang, Phys. Rev. Lett. {\bf111},
    136804 (2013).
\bibitem{kitagawa11}T. Kitagawa, T. Oka, A. Brataas, L. Fu, and E Demler, Phys.
    Rev. B {\bf84}, 235108 (2011).
\bibitem{qiao14b}Z. Qiao, W. Ren, H. Chen, L. Bellaiche, Z. Zhang, A. H.
    MacDonald, and Q. Niu, Phys. Rev. Lett. {\bf112}, 116404 (2014).
\bibitem{barnea12}T. Pereg-Barnea and G. Refael, Phys. Rev. B {\bf85}, 075127
    (2012).
\bibitem{wright13}A. R. Wright, Sci. Rep. {\bf3}, 2736 (2013).
\bibitem{zhang13}X.-L. Zhang, L.-F. Liu, and W.-M. Liu, Sci. Rep. {\bf3}, 2908
    (2013).
\bibitem{kaloni14}T. P. Kaloni, N. Singh, and U. Schwingenschl\"{o}gl, Phys.
    Rev. B {\bf89}, 035409 (2014).
\bibitem{xiao07}D. Xiao, W. Yao, and Q. Niu, Phys. Rev. Lett. {\bf99}, 236809
    (2007).
\bibitem{zhang11}F. Zhang, J. Jung, G. A. Fiete, Q. Niu, and A. H. MacDonald,
    Phys. Rev. Lett, {\bf106}, 156801 (2011).
\bibitem{grujic14b}M. M. Gruji\'{c}, M. \v{Z}. Tadi\'{c}, and F. M. Peeters,
    Phys. Rev. B {\bf90}, 205408 (2014).
\end{thebibliography}
\end{document}